\documentclass[sigconf,nonacm]{acmart}

\AtBeginDocument{%
  }

\usepackage[utf8]{inputenc}
\DeclareUnicodeCharacter{00D7}{\(\times\)}
\usepackage[T1]{fontenc}
\usepackage{booktabs}
\usepackage{tabularx}
\usepackage{array}
\usepackage{enumitem}
\usepackage{amsmath}
\usepackage{multirow}
\usepackage{graphicx}
\usepackage{xcolor}
\usepackage{hyperref}
\usepackage{xspace}
\usepackage{placeins}
\usepackage{caption}
\usepackage{subcaption}

\newcommand{\dataset}{BioMedJImpact\xspace}

\begin{document}

\title[BioMedJImpact: Scientific Impact Analysis of Biomedical Journals]{BioMedJImpact: A Comprehensive Dataset and LLM Pipeline for AI Engagement and Scientific Impact Analysis of Biomedical Journals}

\author{Ruiyu Wang}
\affiliation{%
  \institution{Emory University}
  \city{Atlanta}
  \state{Georgia}
  \country{USA}
}
\email{jonathan.wang@emory.edu}

\author{Yuzhang Xie}
\affiliation{%
  \institution{Emory University}
  \city{Atlanta}
  \state{Georgia}
  \country{USA}
}
\email{yuzhang.xie@emory.edu}

\author{Xiao Hu}
\affiliation{%
  \institution{Emory University}
  \city{Atlanta}
  \state{Georgia}
  \country{USA}
}
\email{xiao.hu@emory.edu}

\author{Carl Yang}
\affiliation{%
  \institution{Emory University}
  \city{Atlanta}
  \state{Georgia}
  \country{USA}
}
\email{j.carlyang@emory.edu}

\author{Jiaying Lu}
\affiliation{%
  \institution{Emory University}
  \city{Atlanta}
  \state{Georgia}
  \country{USA}
}
\email{jiaying.lu@emory.edu}

\renewcommand{\shortauthors}{Wang et al.}

\begin{abstract}
Assessing journal impact is central to scholarly communication, yet existing resources rarely capture how collaboration and artificial intelligence (AI) research jointly shape venue prestige in biomedicine. We present \dataset, a large-scale, biomedical-oriented dataset built from 1.74 million PubMed Central articles across 2,744 journals. \dataset integrates bibliometric indicators, collaboration features, and an LLM-derived AI engagement rate, defined as the proportion of AI-related articles within each journal-year. Specifically, AI engagement rate is extracted through a reproducible three-stage LLM pipeline. We analyze how collaboration intensity and AI engagement rate jointly influence scientific impact across two temporal subsets (2016--2019, 2020--2023). Two main patterns emerge: journals with larger author teams tend to have higher citation impact, while AI engagement rate is positively associated with Impact Factor only in the 2019 subset. To validate the LLM pipeline for deriving the AI engagement rate, we conduct human evaluation, confirming substantial agreement in AI relevance detection and consistent subfield classification. Together, BioMedJImpact provides both a comprehensive dataset at the interface of biomedicine and AI and a validated framework for scalable, content-aware scientometric analysis. Code and dataset are available at \url{https://github.com/JonathanWry/BioMedJImpact}.
\end{abstract}

\begin{CCSXML}
<ccs2012>
 <concept>
  <concept_id>10002951.10003260.10003282</concept_id>
  <concept_desc>Information systems~Data mining</concept_desc>
  <concept_significance>500</concept_significance>
 </concept>
 <concept>
  <concept_id>10010147.10010257.10010293.10010294</concept_id>
  <concept_desc>Computing methodologies~Natural language processing</concept_desc>
  <concept_significance>300</concept_significance>
 </concept>
 <concept>
  <concept_id>10010405.10010489.10010491</concept_id>
  <concept_desc>Applied computing~Life and medical sciences</concept_desc>
  <concept_significance>300</concept_significance>
 </concept>
</ccs2012>
\end{CCSXML}

\ccsdesc[500]{Information systems~Data mining}
\ccsdesc[300]{Computing methodologies~Natural language processing}
\ccsdesc[300]{Applied computing~Life and medical sciences}

\keywords{LLM for feature extraction, journal scientific impact analytics, sequential prompt engineering}

\maketitle

\section{Introduction}
The scientific impact of journals plays a central role in academic communication, influencing research visibility, funding allocation, and institutional evaluation~\cite{garfield1955citation}. In biomedicine, metrics such as the impact factor (IF), citation counts, and journal ranking (e.g., JCR quartiles, SCImago Journal Rank) serve as key decision-making tools for authors, institutions, and funders~\cite{dong2005_if_revisited}. These indicators are sensitive to changes in the research landscape. For example, many general medical journals such as \textit{The New England Journal of Medicine} and \textit{The Lancet} saw sharp spikes in IF in 2021 due to the COVID-19 publication surge, followed by a return toward pre-pandemic levels in 2022~\cite{kim2024_covid_if}. Meanwhile, artificial intelligence (AI) has increasingly transformed biomedical research~\cite{reason2024_pico_genai,xie2025kerap}, reshaping how research is conducted and evaluated, and potentially altering traditional scientific impact indicators.

Existing open datasets on scientific impact (e.g., AMiner~\cite{tang2008_aminer}, DBLP~\cite{ley2002_dblp}, and Microsoft Academic Graph~\cite{wang2020_mag}) have greatly advanced large-scale analyses of scholarly networks and citation behavior; DBLP focuses on computer science, while AMiner and MAG provide broad, cross-disciplinary coverage. However, these existing resources are not designed specifically for the biomedical domain, and they lack the granularity needed to capture AI’s influence within biomedical research. 
To tackle these problems, we leverage multi-source data to build a new dataset on biomedical journals' scientific impact, called \dataset. \dataset\ integrates three major categories of features:
(a) Bibliometric indicators (e.g., impact metrics, citation counts),
(b) Collaboration indicators (e.g., author diversity, institutional diversity), and
(c) AI-related indicators (e.g., AI engagement rate, AI subfield distribution) that quantify the presence and distribution of AI-related research across journals. Specifically, bibliometric indicators are sourced from the Journal Citation Reports (JCR) and CiteFactor; collaboration indicators are derived from the PubMed Central Open Access subset (PMC-OA) metadata, capturing author and institutional structures; and the AI-related indicators are derived from article abstracts using a large language model (LLM)-based pipeline, enabling us to identify AI-related publications and their associated subfields. For downstream analysis, we construct two temporal subsets:
\textbf{BioMedJImpact-2019} (2016--2019) and
\textbf{BioMedJImpact-2023} (2020--2023).

In total, we build \dataset, a comprehensive dataset for analyzing biomedical journals' scientific impact, consisting of 2,744 journals. We derive 55 features covering bibliometric, collaboration, and AI-related indicators. Using PMC-OA, we matched journals by source and year and identified 1,740,112 papers, processed through our LLM-based pipeline, yielding an overall AI engagement rate of 3.77\%. Collaboration intensity—especially larger and more diverse author teams—is consistently associated with higher citation impact, while AI engagement rate is associated with Impact Factor only in BioMedJImpact-2019. Overall, \textbf{\dataset} combines LLM-derived semantic indicators with traditional bibliometric and collaboration features to support scalable analysis of how content and collaboration shape scientific impact.
\enlargethispage{2\baselineskip}

\vspace{-0.1cm}
\section{Related Work}
\vspace{-0.1cm}
\subsection{Scientific Impact Modeling}
% \vspace{-0.1cm}
% please refer to https://dl-acm-org.proxy.library.emory.edu/doi/pdf/10.1145/2684822.2685314

Scientific impact has long been a core concern in scientometrics and information science, offering insight into how scholarly influence accumulates and providing practical tools for evaluating research quality, allocating funding, and guiding publication strategies~\cite{garfield1955citation}. Among various indicators, citations remain the primary quantitative signal, forming the basis of a family of \emph{citation-based indicators} used to assess journals, authors, and individual papers~\cite{bornmann2008_citationreview}. At the journal level, Garfield’s journal impact factor formalized citation aggregation as a venue-level indicator~\cite{garfield2006_jifmeaning}, while the JCR quartile scheme (Q1--Q4) situates journals within disciplinary hierarchies, providing a coarse yet actionable measure of prestige for authors, editors, and institutions~\cite{krampl2019jcr}. Despite well-documented limitations including field-normalization challenges and citation skewness, these citation-based indicators remain interpretable benchmarks that correlate with long-term scientific attention~\cite{wang2013_science}.

Beyond citation-based indicators, research has explored how various factors contribute to scientific impact. Collaborative indicators such as team size and co-authorship networks have been shown to correlate with citation influence across disciplines \cite{lariviere2015_teamsize}. Studies also indicate that thematic and linguistic content learned from titles and abstracts encodes meaningful cues of scholarly influence \cite{cohan2020_specter}. 
\vspace{-8pt}
\subsection{Large Language Model-Based Feature Extraction}
% \vspace{-2pt}

Alongside traditional feature extraction methods, recent advances in LLMs have fundamentally changed how features can be mined from scientific papers. Conventional feature extraction pipelines often rely on task-specific supervised models for named-entity, relation, and event extraction~\cite{wadden2019_entity_event}. By contrast, LLMs enable prompt-based extraction that can screen documents for topical relevance, identify domain-specific terms, and map those terms to controlled taxonomies. Recent surveys document strong zero-shot and few-shot performance of LLMs for generative information extraction across domain-general tasks including named-entity, relation, and event extraction~\cite{xu2024_llm_generative_ie}.

In biomedical corpora, LLMs have been applied to instruction following information extraction across core tasks including named–entity recognition, relation extraction, and procedure extraction~\cite{xie2025hypkg,bhasuran2025preliminary}. In concrete biomedical applications, LLMs have likewise demonstrated practical utility. In radiology, the RadEx framework provides a modular architecture for developing information-extraction systems from free-text radiology reports, supporting both generative and encoder-based models~\cite{reichenpfader2026_radex}. Similarly, LLMs have demonstrated their feasibility, accuracy, and efficiency for large-scale study design elements (PICO) extraction from clinical abstracts in PubMed~\cite{reason2024_pico_genai}.

\section{Dataset Construction}
\vspace{-0.5pt}
\subsection{Multisource Integration for Initial Dataset Construction}
\vspace{-0.5pt}
In this study, we curate \dataset, a comprehensive journal-level dataset for large-scale analysis of biomedical journal impact and AI engagement, by integrating data from:
(i) the \textit{PubMed Central Open Access subset} (PMC-OA)~\cite{PMC-OA} for full-text and metadata of biomedical articles,
(ii) \textit{Journal Citation Reports} (JCR)~\cite{krampl2019jcr} for journal bibliometric records including historical impact metrics and citation information, and 
(iii) the \textit{Directory of Open Access Journals} (DOAJ)~\cite{morrison2017doaj} for journal-level open-access policies and publication practices.
Based on these resources, we assemble 17 per–journal, per–year indicators (see Table~\ref{tab:feature-summary} for details). Among all data sources, PMC-OA serves as the core foundation of \dataset. It provides full-text and metadata for \textbf{4,298} biomedical journals. After matching these journals with available bibliometric records from JCR, we retain \textbf{2,744} journals for downstream analysis of content, citation patterns, and collaboration indicators. Of these, \textbf{1,694} journals are indexed in the DOAJ, enabling the integration of open-access policies and publication practices into the dataset.
% The finalized version of \dataset\ will be released to the research community upon acceptance to promote transparency, reproducibility, and further investigation into biomedical journal impact.

\vspace{-0.1cm}
\begin{table}[htbp!]
\centering
\small
\caption{Summary of statistics for the \dataset-\textbf{2019} and \dataset-\textbf{2023} subsets.} 
\vspace{-0.1cm}
%The “Target Year” refers to the year in which Impact Factor (IF), Quartile, and three-year citation counts (\emph{Total Cites\_3Y}) are used as outcome variables.
%\emph{Std} denotes standard deviation.}
\label{tab:dataset-summary}
\begin{tabular}{lcc}
\toprule
Statistic \textbackslash{} Sub Dataset & -2019 & -2023 \\
\midrule
\# Journals & 1367 & 2685 \\
\hline
\# Journals with Quartile & 1243 & 2321 \\
\quad Percentage of Q1 journals & 57.2\% & 46.92\% \\
\hline
\# Journals with IF & 1367 & 2685 \\
\quad Avg IF & 3.43 & 3.35\\
\quad Std IF & 3.12 & 4.01 \\
\hline
\# Journals with Total Cites\_3Y & 1247 & 2314 \\
\quad Avg Total Cites (3Y) & 27263 & 33018\\
\quad Std Total Cites (3Y) & 116497 & 132542 \\
\bottomrule
\end{tabular}
\vspace{-0.3cm}
\end{table}

\subsubsection{Dataset/Year Split.} 
To facilitate downstream modeling and isolate temporal effects, we partition the unified dataset into two temporal subsets: \textbf{BioMedJImpact-2019} (2016–2019) and \textbf{-2023} (2020–2023). Within each subset, we focus on three commonly used journal-level targets: \emph{Impact Factor}, \emph{Quartile}, and \emph{Total Cites (3Y)}. These targets are widely used in academic assessment systems: Impact Factor reflects short-term citation influence, Quartile indicates a journal’s relative standing within its subject category, and Total Cites (3Y) captures sustained citation accumulation. The temporal split accounts for structural shifts in publishing and citation dynamics during the COVID-19 period, reducing confounding from pandemic-related disruptions. For each subset, we retain only journals with valid IFs in the subset’s target year. After filtering, \textbf{BioMedJImpact-2019} contains 1{,}367 journals and \textbf{BioMedJImpact-2023} contains 2{,}685 journals, with over 90\% of journals including quartile rankings and citation-based metrics. Table~\ref{tab:dataset-summary} summarizes the retained sets.
\subsubsection{Bibliometric Indicators.}
\dataset integrates bibliometric indicators from multiple publicly available sources. Historical journal indicators are collected from JCR hosted on ResearchGate\footnote{\href{https://www.researchgate.net/}{https://www.researchgate.net/}} (2016–2024), and missing values were supplemented using CiteFactor\footnote{\href{https://www.citefactor.org}{https://www.citefactor.org}}. Extracted fields include journal title, ISSN/EISSN, subject category, quartile ranking (Q1--Q4), impact factor, and total citations. Policy attributes are integrated via cross-referencing with the DOAJ~\cite{morrison2017doaj}, which provides publication delay (in weeks), author copyright-retention status, and article processing charges. All sources are harmonized by ISSN/EISSN as unique identifiers, with fuzzy title matching applied for unresolved cases.
\vspace{-5pt}
\subsubsection{Collaboration Indicators.}
To characterize author collaboration patterns, we process full-text XML archives from PMC-OA~\cite{PMC-OA}. PMC-OA provides rich article-level metadata, including author affiliations, article types, and author-supplied keywords. From each article, we extract the number of distinct institutions and participating countries using both structured tags (\texttt{<institution>}, \texttt{<country>}) and fallback string-pattern matching when such tags are absent. 
 
We additionally define a \emph{cross-country collaboration rate} as the proportion of articles with author affiliations spanning multiple countries. These metrics enable standardized comparisons of institutional and international collaboration intensity across disciplines and temporal spans.
\vspace{-5pt}
\begin{figure*}[h!]
\vspace{-0.5cm}
\centering
\begin{tabular}{cc}
    % -------- Row 1 --------
    \includegraphics[width=0.35\textwidth]{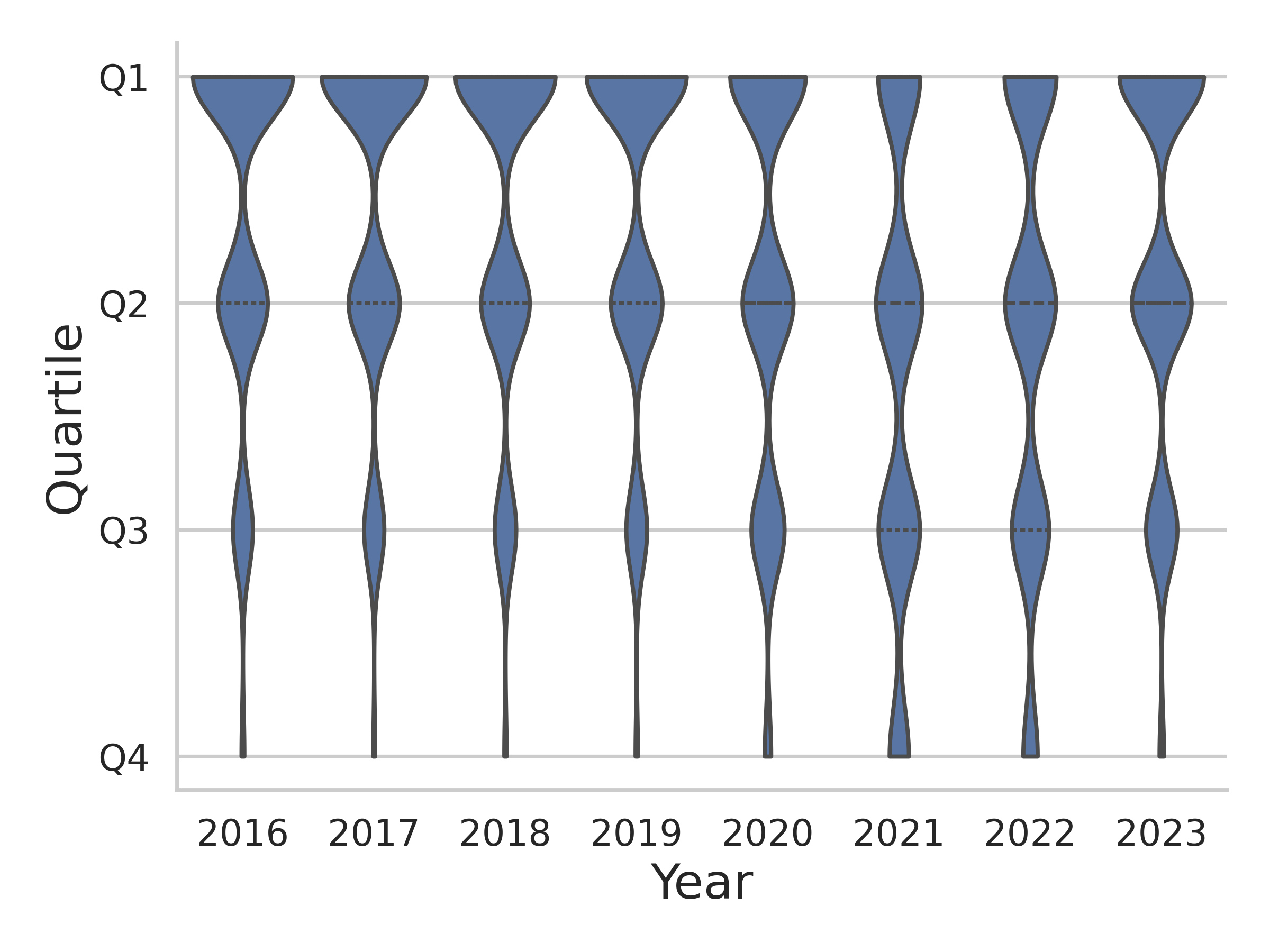} &
    \includegraphics[width=0.35\textwidth]{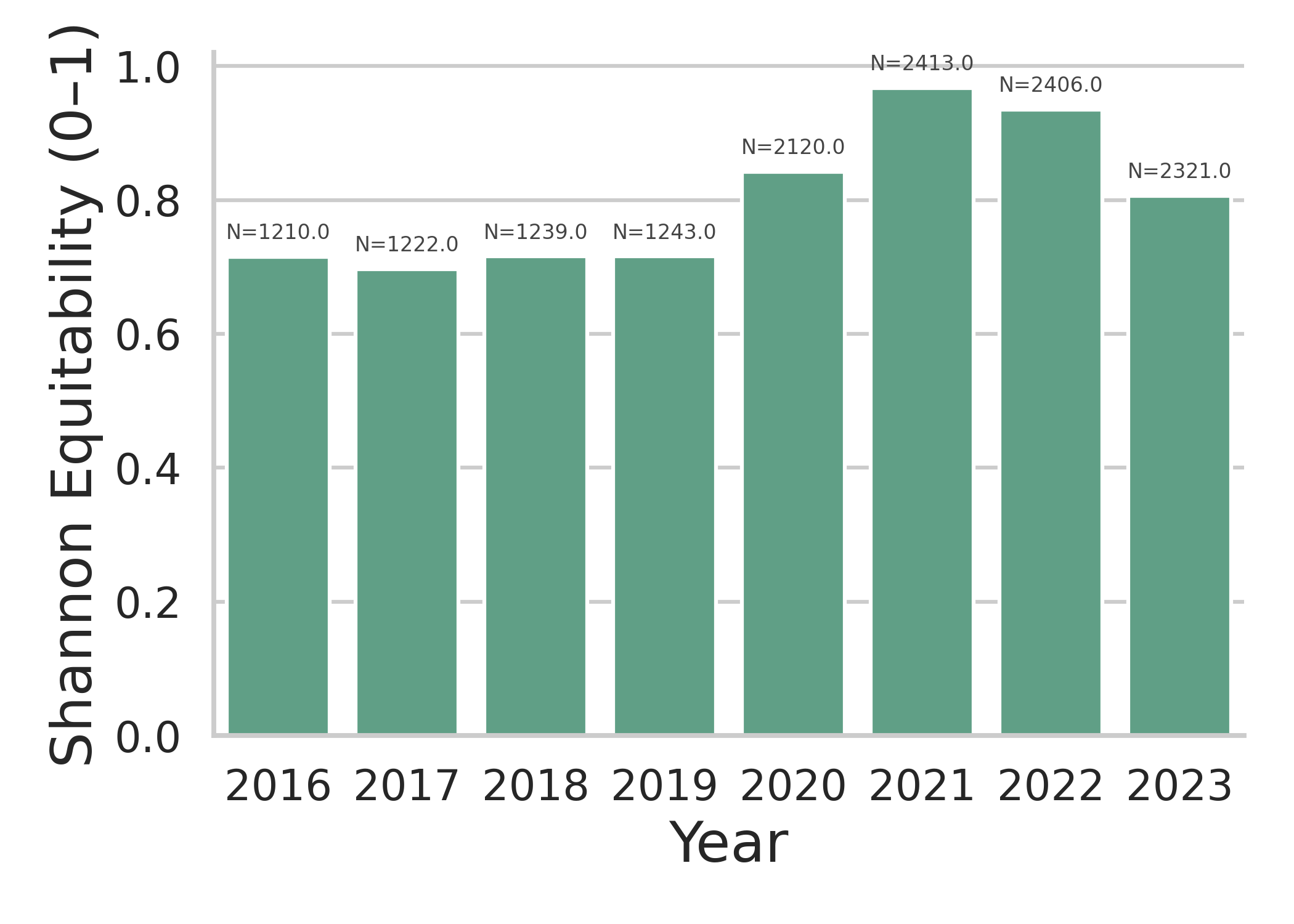} \\[-2pt]
    (a) Quartile distribution & (b) Quartile balance  \\[6pt]

    % -------- Row 2 --------
    \includegraphics[width=0.35\textwidth]{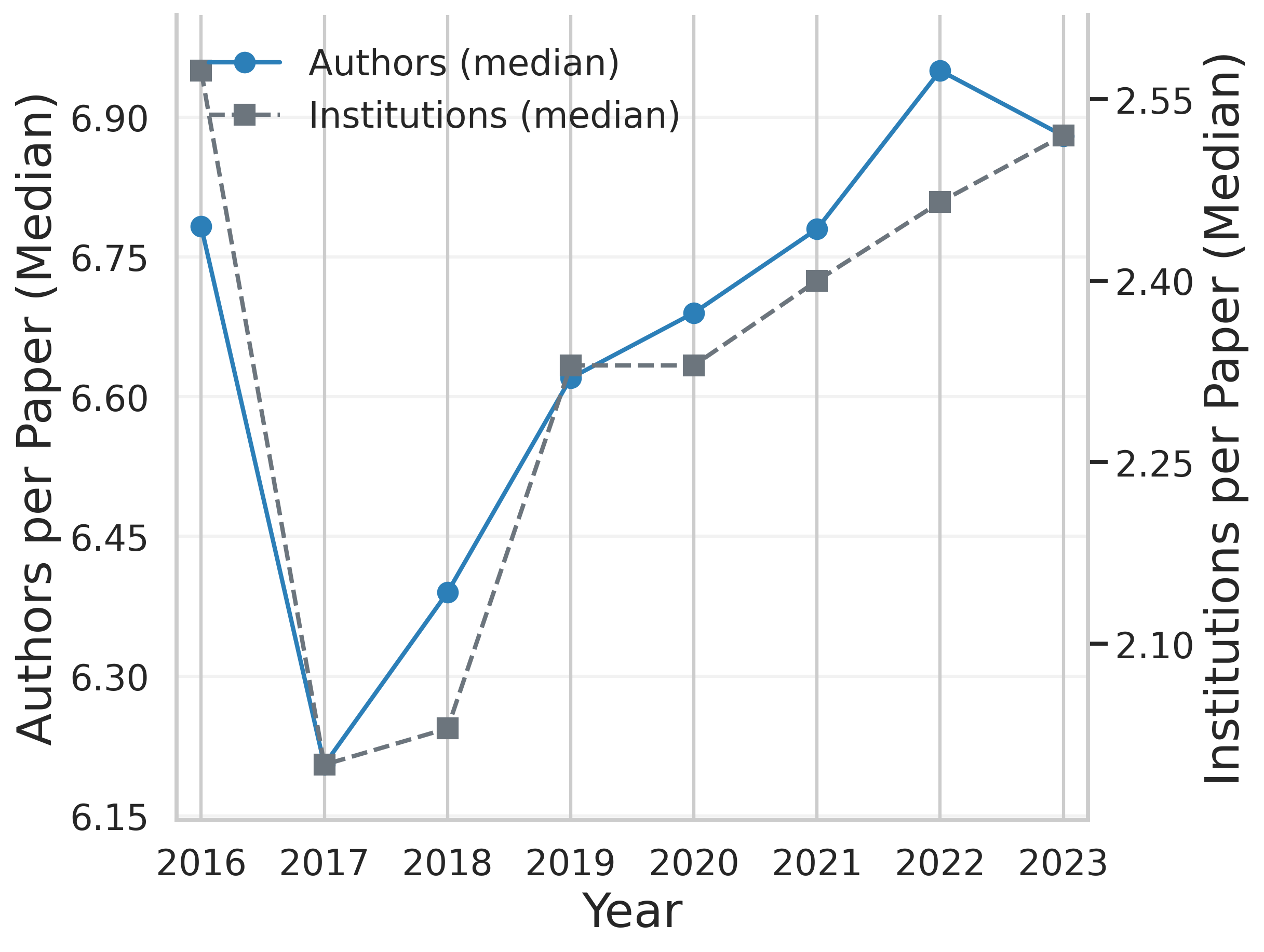} &
    \includegraphics[width=0.35\textwidth]{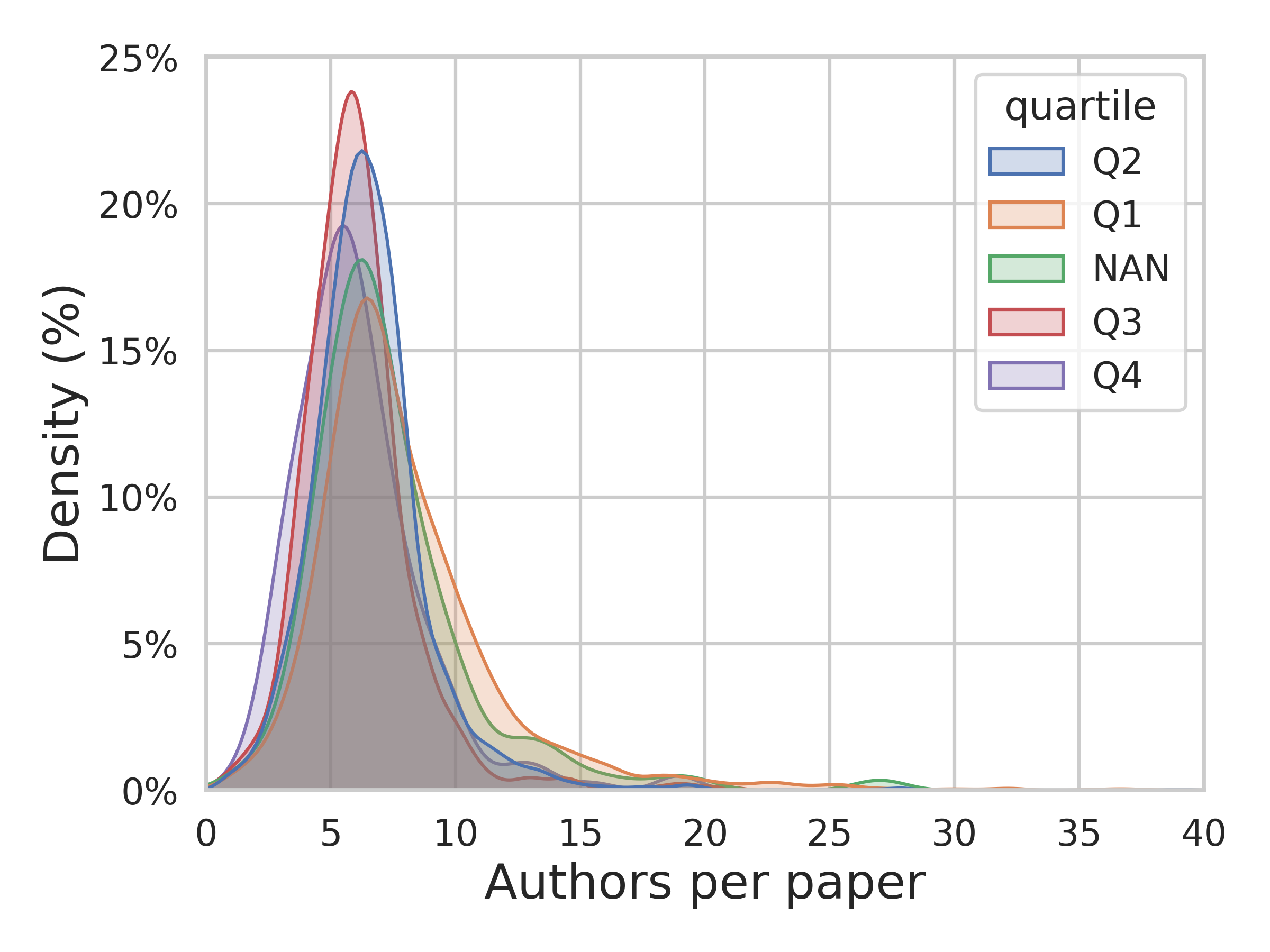} \\[-2pt]
    (c) Collaboration trends & (d) Authors per paper by quartile
\end{tabular}
\caption{Dataset overview of quartile and collaboration indicators over time.}
\label{fig:quartile-collaboration-grid}
\end{figure*}
\subsubsection{Dataset Overview.}  
After deriving the bibliometric and collaboration indicators, we conduct dataset characteristic analytics.
Figure~\ref{fig:quartile-collaboration-grid}(a,b) depict the temporal distribution of journal quartiles from 2016 to 2023. The overall quartile composition remains stable, with the proportion of Q1 journals exhibiting a modest decrease between 2021 and 2022 (also reflected in the increased Shannon equitability index).
Figure~\ref{fig:quartile-collaboration-grid}(c,d) show longitudinal trends in author and institutional collaboration. The median number of authors per paper ranges from 6.2 to 6.9, and no substantial differences in collaboration patterns are observed across Q1–Q4 journals.

% Sec 3.2
% merged into data construction
\vspace{-0.6\baselineskip} 
\subsection{LLM-Based Feature Enrichment for Journal AI Engagement}

To capture how journals engage with the rapidly growing field of AI, we extend \dataset beyond traditional bibliometric and collaboration indicators by incorporating features extracted through an LLM-based pipeline. These content-level signals quantify each journal’s AI engagement and, when combined with structural and collaboration features, yield a more comprehensive dataset. The pipeline produces AI-related article shares, which are integrated into the full feature set detailed in Table \ref{tab:feature-summary}.

\begin{table}[ht!]
\vspace{-0.1cm} 
\caption{Summary of features. \textnormal{Feature \textsubscript{Y-1}, \textsubscript{Y-2}, and \textsubscript{Y-3} denote covariates from one to three years prior to the prediction year.}}
\vspace{-0.2cm}
\label{tab:feature-summary}
\centering
\small
\resizebox{\columnwidth}{!}{%
\begin{tabular}{p{0.4\columnwidth} p{0.6\columnwidth}}
\toprule
\textbf{Feature Group} & \textbf{Features Included} \\
\midrule
\textbf{Bibliometric Indicators} 
& \textbullet\ Impact Factor \textsubscript{Y-1}, \textsubscript{Y-2}, \textsubscript{Y-3} \newline
% \textbullet\ Category \newline
\textbullet\ Quartile \textsubscript{Y-1}, \textsubscript{Y-2}, \textsubscript{Y-3} \newline
\textbullet\ Total Cites (3Y) \textsubscript{Y-1}, \textsubscript{Y-2}, \textsubscript{Y-3} \newline
\textbullet\ Total References \textsubscript{Y-1}, \textsubscript{Y-2}, \textsubscript{Y-3} \newline
\textbullet\ Publication Count \textsubscript{Y-1}, \textsubscript{Y-2}, \textsubscript{Y-3}  \newline
\textbullet\ Publication Delay (in weeks) \newline
\textbullet\ Author Copyright Retention \newline
\textbullet\ Article Processing Charges \newline
\textbullet\ Subject Category \\ 
\midrule
\textbf{Collaboration Indicators} 
& \textbullet\ Avg. Authors \textsubscript{Y-1}, \textsubscript{Y-2}, \textsubscript{Y-3} \newline
\textbullet\ Std. Authors \textsubscript{Y-1}, \textsubscript{Y-2}, \textsubscript{Y-3} \newline
\textbullet\ Author Quartiles \textsubscript{Q25, Q50, Q75 × Y-1, Y-2, Y-3} \newline
\textbullet\ Avg. Institutions \textsubscript{Y-1}, \textsubscript{Y-2}, \textsubscript{Y-3} \newline
\textbullet\ Std. Institutions \textsubscript{Y-1}, \textsubscript{Y-2}, \textsubscript{Y-3} \newline
\textbullet\ Institution Quartiles \textsubscript{Q25, Q50, Q75 × Y-1, Y-2, Y-3} \newline
\textbullet\ Cross-country collaboration rate \textsubscript{Y-1}, \textsubscript{Y-2}, \textsubscript{Y-3} \\
\midrule
\textbf{AI-Related Indicators} 
& \textbullet\ AI Engagement rate \textsubscript{Y-1}, \textsubscript{Y-2}, \textsubscript{Y-3} \\
\bottomrule
\end{tabular}
}
\vspace{-0.4cm}
\end{table}
\subsubsection{LLM-Based Content Analysis on AI}
\begin{figure}[t]
  \centering
  \vspace{-0.2cm} 
  \includegraphics[width=0.45\textwidth]{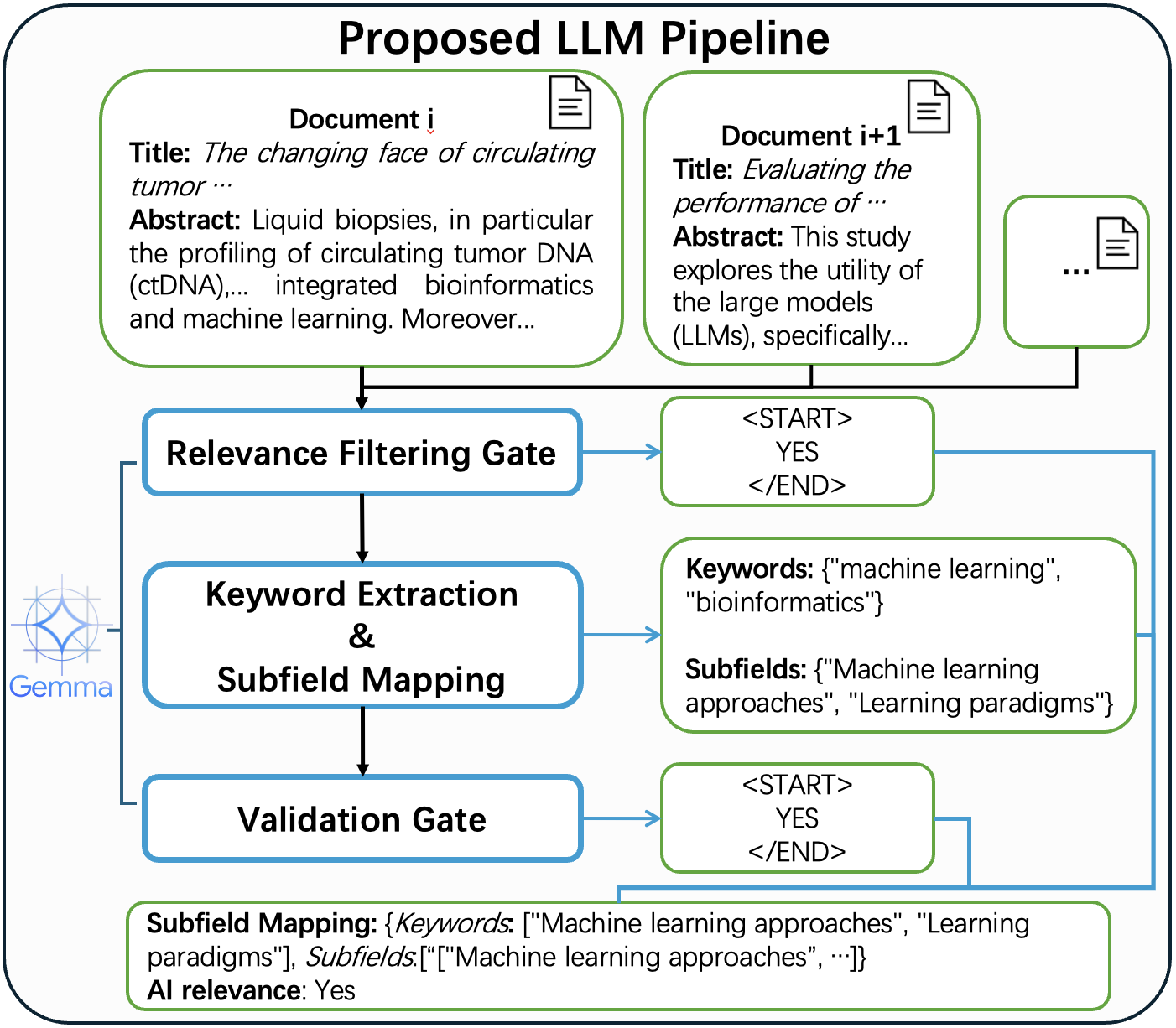}
  \caption{Overview of our LLM pipeline for AI engagement rate analysis. \textnormal{Step~1 determines whether the input document is AI-relevant. Step~2 extracts AI terms and maps them to a taxonomy. Step~3 further reduces ambiguity and false positives.}}
  \label{fig:llm-pipeline}
\vspace{-0.4cm}
\end{figure}

To quantify AI involvement, we annotate PMC-OA article abstracts with an LLM to estimate journal-level AI engagement (Figure~\ref{fig:llm-pipeline}). We implement a three-step LLM pipeline with \textit{Gemma-3-12B}~\cite{gemma32025technical}: 
\begin{enumerate}
  \item \textbf{Relevance Filtering Gate:} Each abstract is screened by an LLM prompt to determine whether it is relevant to artificial intelligence or machine learning. Abstracts containing phrases such as “deep learning–based model,” “AI-assisted diagnosis,” or “neural network training” are labeled as AI-relevant. Non-technical mentions (e.g., “intelligent design”) are filtered out.
  
  \item \textbf{Keyword Extraction and Subfield Mapping:} Abstracts identified as AI-relevant are then processed by a second LLM prompt that simultaneously (1) extracts AI-related keywords (e.g., CNN, reinforcement learning) and (2) maps each abstract to one or more predefined AI subfields, including \emph{Natural Language Processing}, \emph{Computer Vision}, \emph{Learning Algorithms}, \emph{Knowledge Representation}, \emph{Search methodologies}, and \emph{Distributed AI}. This integrated keyword–subfield reasoning step enables consistent subfield assignment and supports downstream analysis of AI research themes.
  
  \item \textbf{Validation Gate:} A secondary verification prompt re-evaluates extracted keywords to confirm their alignment with AI subfields and removes ambiguous or noisy terms (e.g., “training session”). This ensures semantic precision and minimizes false positives in downstream statistical analyses.
\end{enumerate}

\begin{figure}[t]
\centering
\vspace{-0.3cm}
\includegraphics[width=0.8\columnwidth]{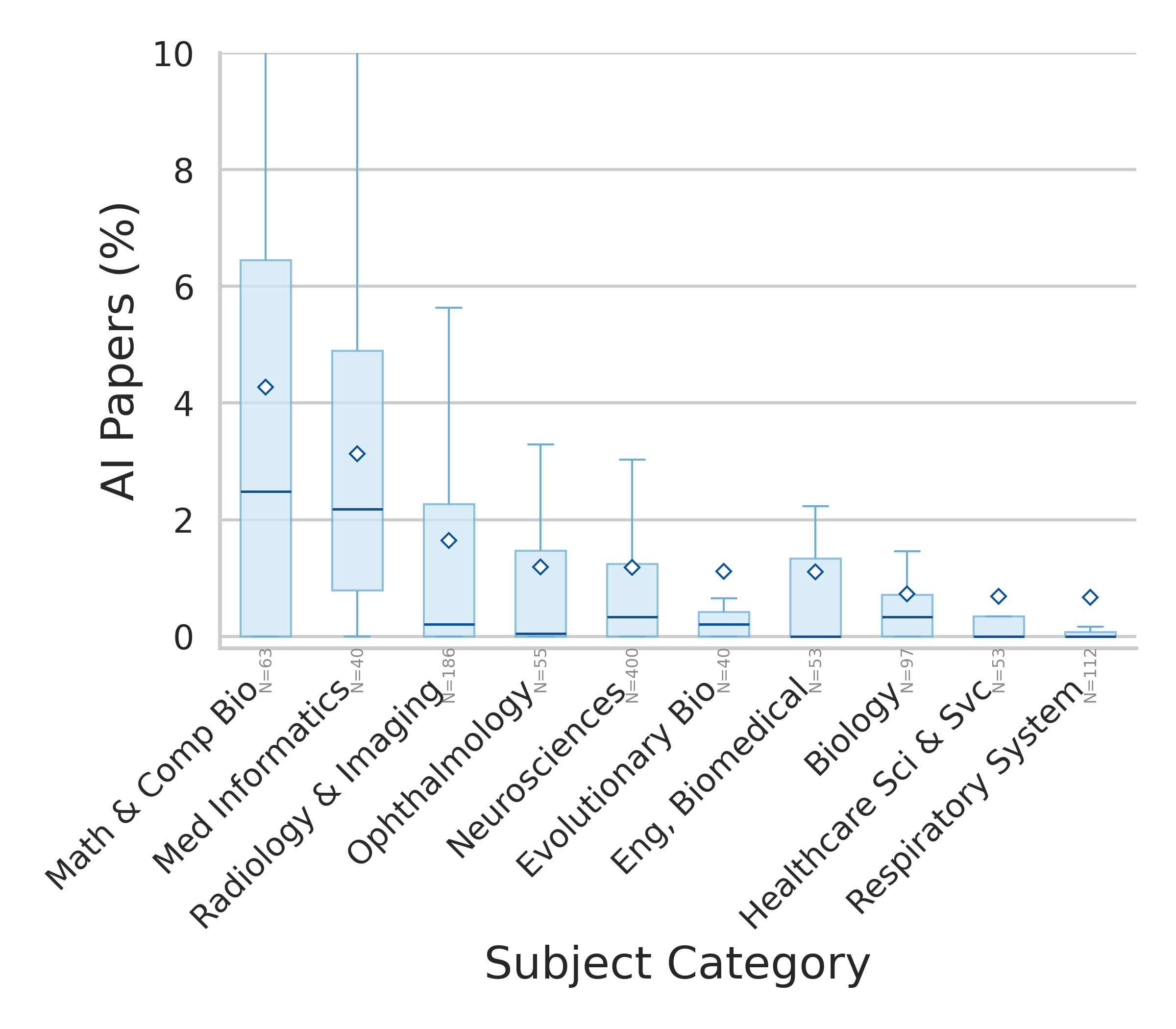}
\vspace{-0.10cm}

{\small \centering (a) AI\% by subject category\par}

\vspace{0.35cm}
\includegraphics[width=0.8\columnwidth]{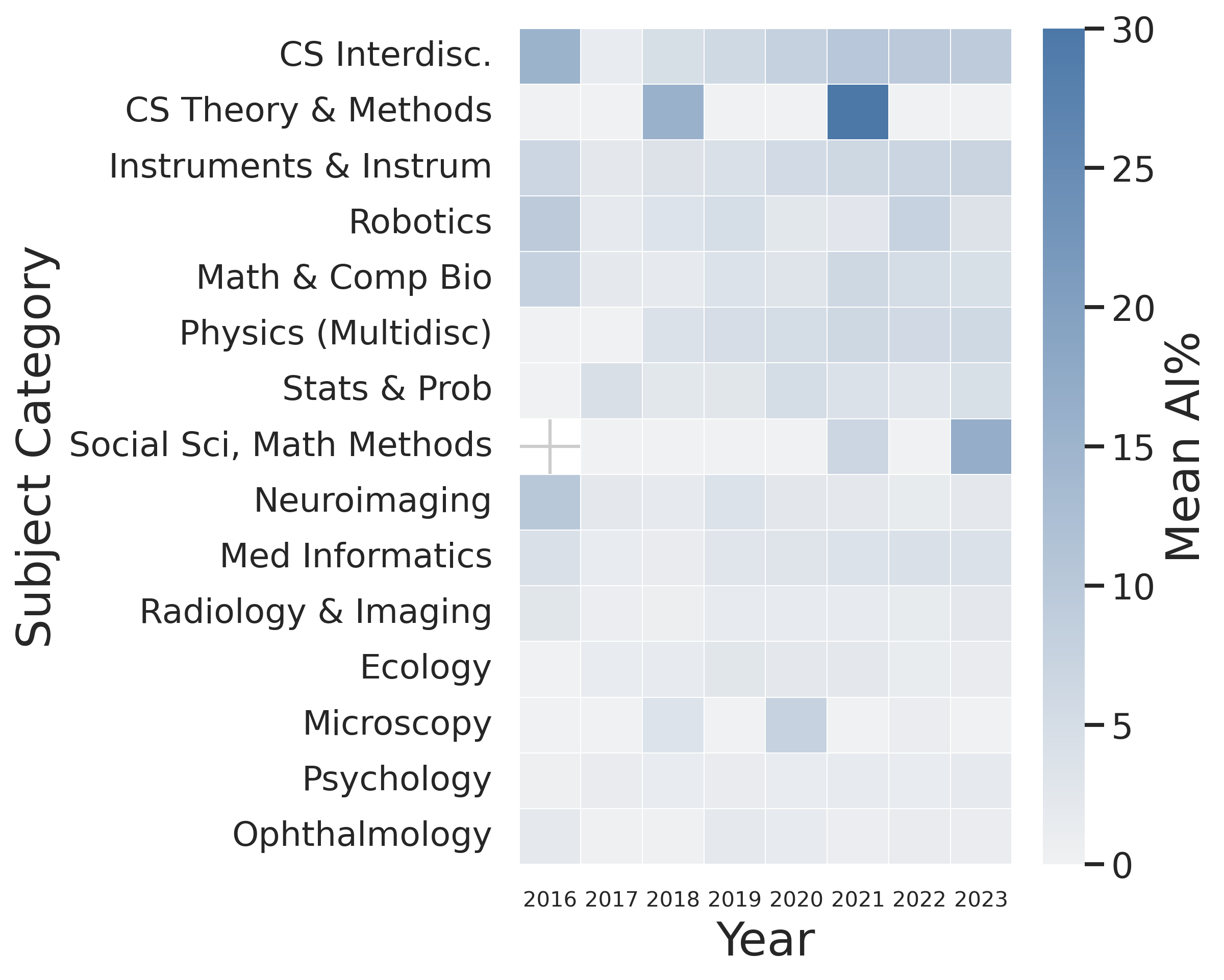}

\vspace{-0.1cm}
{\small \centering (b) AI\% by year and subject category}

\vspace{-0.1cm}
\caption{AI engagement rate patterns, where the rate denotes the proportion of AI-related articles within each journal–year. \textnormal{Panel (a): Top-10 subject categories ranked by pooled mean AI\% across all journal--year records; boxes show distributions, diamonds show means. Panel (b): Top-15 categories ranked by year-normalized mean AI\%---the mean across journals per category--year, then averaged over years.}}
\label{fig:ai-engagement-grid}
\vspace{-0.7cm}
\end{figure}

\noindent From this pipeline, we derive the \emph{AI engagement rate}, defined as the proportion of articles within each journal--year that are classified as AI-related:
\(E_{j,t} = N^{\text{AI}}_{j,t} / N^{\text{total}}_{j,t}\),  where \(N^{\text{AI}}_{j,t}\) is the number of AI-related abstracts in journal \(j\) during year \(t\), and \(N^{\text{total}}_{j,t}\) is the total number of abstracts published by that journal in the same year.
\vspace{-0.2cm}
\subsubsection{AI Engagement Patterns.}
We summarize AI-related publication trends in Figure~\ref{fig:ai-engagement-grid}, based on LLM-annotated journal content. AI activity is most prominent in multidisciplinary areas linked to computer science and mathematics, with additional strong engagement in imaging-focused fields such as \emph{Radiology and Medical Imaging}, and \emph{Neuroscience}. 
In terms of dataset coverage, AI engagement indicators are available for 902 journals in 2019, including 631 with full three-year AI coverage. By 2023, this expands to 1{,}010 journals, with 803 fully covered over three years. Collaboration features show similarly broad availability, present in 1{,}095 journals in 2019 and 979 in 2023.
\begin{table*}[!t]
\vspace{-0.2cm}
\caption{LME/OLS correlation analysis for \textbf{\dataset-2019}. \textnormal{Reported significance uses Bonferroni-adjusted p-values within each model ($*$ denotes $p_{\mathrm{adj}}<0.05$, $**$ denotes $p_{\mathrm{adj}}<0.01$, and $***$ denotes $p_{\mathrm{adj}}<0.001$).}}
\centering
\small
\setlength{\tabcolsep}{7pt}
\begin{tabular}{l l r r c c}
\toprule
\textbf{Target} & \textbf{Variable} & \textbf{Coef.} & \textbf{Std. Err.} & \textbf{95\% CI (L--H)} & \textbf{Signif.} \\
\midrule
\multirow{2}{*}{Impact Factor}
 & Avg\_Authors\_2016         & 0.882  & 0.092  & [0.701, 1.063] & *** \\
 & AI\_Engagement\_Rate\_2018 & 19.38  & 4.84   & [9.89, 28.88]  & **  \\
\midrule
\multirow{6}{*}{Total Cites}
 & Publication\_Count\_2016   & -146.31 & 15.12 & [-175.94, -116.68] & *** \\
 & Publication\_Count\_2018   & 136.93  & 14.31 & [108.89, 164.97]   & *** \\
 & Publication\_Count\_2017   & 108.13  & 19.45 & [70.00, 146.25]    & *** \\
 & Total\_Refs\_2018          & -0.775  & 0.185 & [-1.137, -0.413]   & *** \\
 & Author\_Copyright\_Retention & -30673.59 & 7468.64 & [-45311.86, -16035.31] & ** \\
 & Std\_Institutions\_2017    & 10995.30 & 3305.07 & [4517.48, 17473.12] & * \\
\bottomrule
\end{tabular}
\vspace{-0.25cm}
\label{tab:reg_j1}
\end{table*}
\section{Correlation Analysis}
\label{sec:dataset-correlation}
To examine how journal features shape biomedical scientific impact, we fit linear mixed-effects (LME) models~\cite{pinheiro2000mixed} with subject-category random intercepts, which account for persistent disciplinary differences while estimating within-field associations. In practice, the random-effect variance occasionally collapses to approximately zero or becomes non-identifiable. In such cases, we refit the same specification using ordinary least squares (OLS), retaining the identical set of fixed effects. 

Because each model includes multiple fixed-effect coefficients, we correct for multiple comparisons \emph{within each fitted model} using a Bonferroni adjustment over all fixed effects (excluding the intercept). Unless otherwise stated, statistical significance throughout this section refers to Bonferroni-adjusted p-values \cite{dunn1961multiple}  ($p_{\mathrm{adj}}$), with \(*\,p_{\mathrm{adj}}<0.05\), \(**\,p_{\mathrm{adj}}<0.01\), and \(***\,p_{\mathrm{adj}}<0.001\). Tables~\ref{tab:reg_j1} and \ref{tab:reg_j2} report only coefficients that remain significant after this correction.

\begin{table*}[!t]
\vspace{-0.2cm}
\caption{LME/OLS correlation analysis for \textbf{\dataset-2023}. \textnormal{Reported significance uses Bonferroni-adjusted p-values within each model ($*$ denotes $p_{\mathrm{adj}}<0.05$, $**$ denotes $p_{\mathrm{adj}}<0.01$, and $***$ denotes $p_{\mathrm{adj}}<0.001$).}}
\small
\centering
\setlength{\tabcolsep}{6pt}
\begin{tabular}{l l r r c c}
\toprule
\textbf{Target} & \textbf{Variable} & \textbf{Coef.} & \textbf{Std. Err.} & \textbf{95\% CI (L--H)} & \textbf{Signif.} \\
\midrule
\multirow{1}{*}{Impact Factor}
 & Avg\_Authors\_2021         & 0.534  & 0.119  & [0.300, 0.767] & *** \\
\midrule
\multirow{5}{*}{Total Cites}
 & Publication\_Count\_2022   & 114.31  & 21.43  & [72.31, 156.32] & *** \\
 & Author\_Copyright\_Retention & -55034.58 & 12306.28 & [-79154.44, -30914.72] & *** \\
 & Std\_Authors\_2020         & 3369.13 & 769.05  & [1861.82, 4876.44] & *** \\
 & Std\_Authors\_2022         & 3018.81 & 880.83  & [1292.42, 4745.20] & * \\
 & Std\_Authors\_2021         & -3672.30 & 1164.23 & [-5954.14, -1390.45] & * \\
\bottomrule
\end{tabular}%

\label{tab:reg_j2}
\end{table*}

For \textbf{Impact Factor}, collaboration intensity remains positively associated after multiple-testing correction. In \dataset-2019, the average number of authors per paper (\texttt{Avg\_Authors\_2016}) is strongly associated with higher impact (\(\beta=0.882\), \(p_{\mathrm{adj}}<0.001\)). AI engagement is also significant in \dataset-2019: a higher AI publication share in 2018 (\texttt{AI\_Engagement\_Rate\_2018}) predicts higher impact (\(\beta=19.38\), \(p_{\mathrm{adj}}<0.01\)). In \dataset-2023, \texttt{Avg\_Authors\_2021} remains positively associated with impact (\(\beta=0.534\), \(p_{\mathrm{adj}}<0.001\)), whereas the corresponding AI engagement variables do not remain significant after Bonferroni correction.
For \textbf{Total Cites}, temporal effects in publication volume persist after correction. In \dataset-2019, publication counts from 2016--2018 and reference volume in 2018 (\texttt{Total\_Refs\_2018}) remain significant, indicating that recent production and referencing practices correlate with subsequent citation totals. Policy and collaboration structure also matter: \textsf{Author\_Copyright\_Retention} is negatively associated with citations (\(\beta=-3.07\times10^{4}\), \(p_{\mathrm{adj}}<0.01\)), while institutional diversity in 2017 (\texttt{Std\_Institutions\_2017}) is positive (\(p_{\mathrm{adj}}<0.05\)). In \dataset-2023, citations are most strongly associated with publication count and authorship variability, while \textsf{Author\_Copyright\_Retention} remains strongly negative.
For \textbf{Quartile}, no fixed-effect coefficient remains significant under the Bonferroni correction in either period, suggesting quartile rank is less robustly explained by the included lagged features than continuous impact outcomes under this conservative correction.

\section{Evaluation on LLM-based Feature Extraction}

\subsection{Quality Evaluation on Select Journal Categories}
\enlargethispage{1\baselineskip}
\begin{figure}[htbp!]
\vspace{-0.1cm}
  \centering \includegraphics[width=0.16\textwidth]{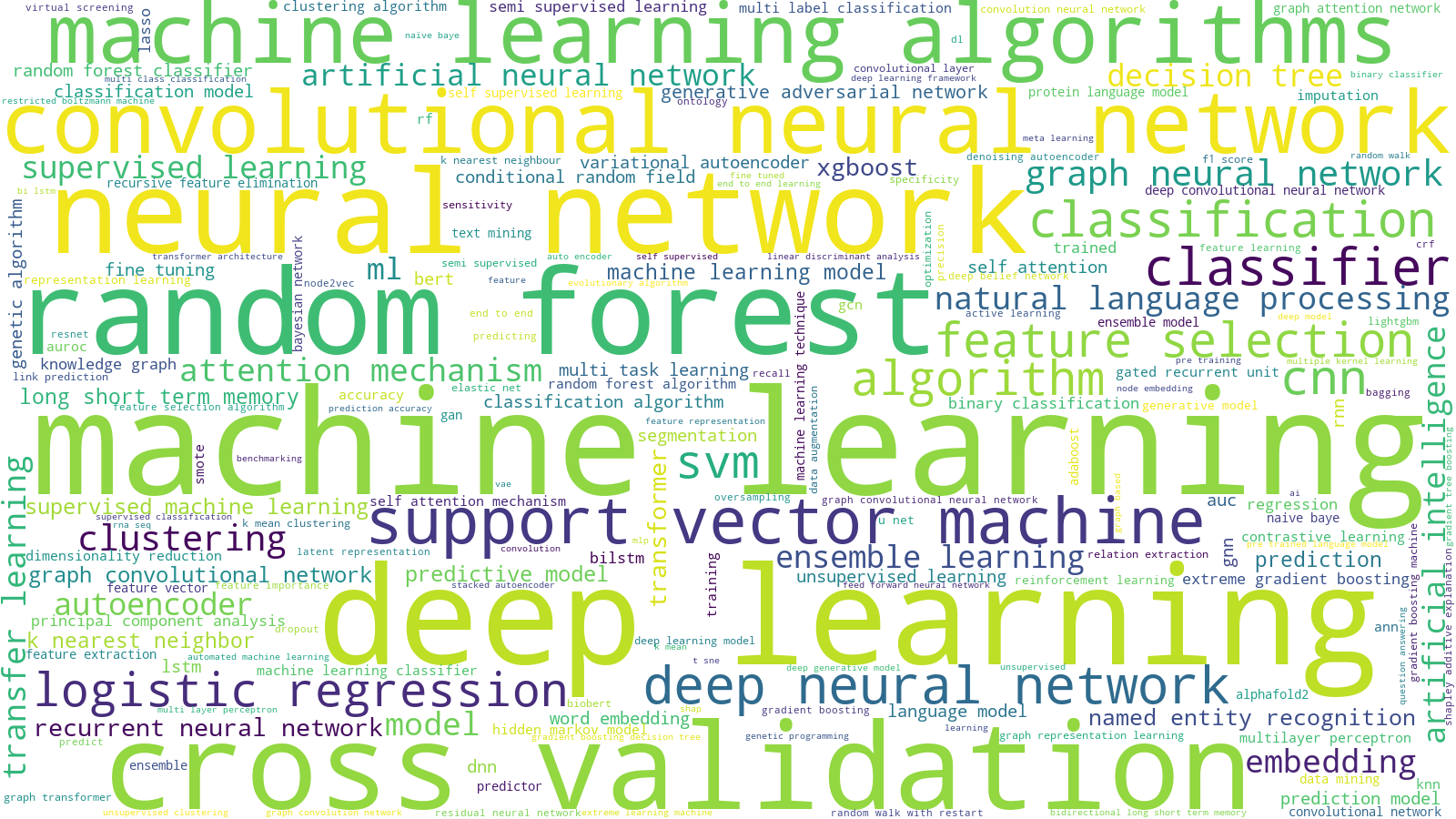}
  \includegraphics[width=0.15\textwidth]{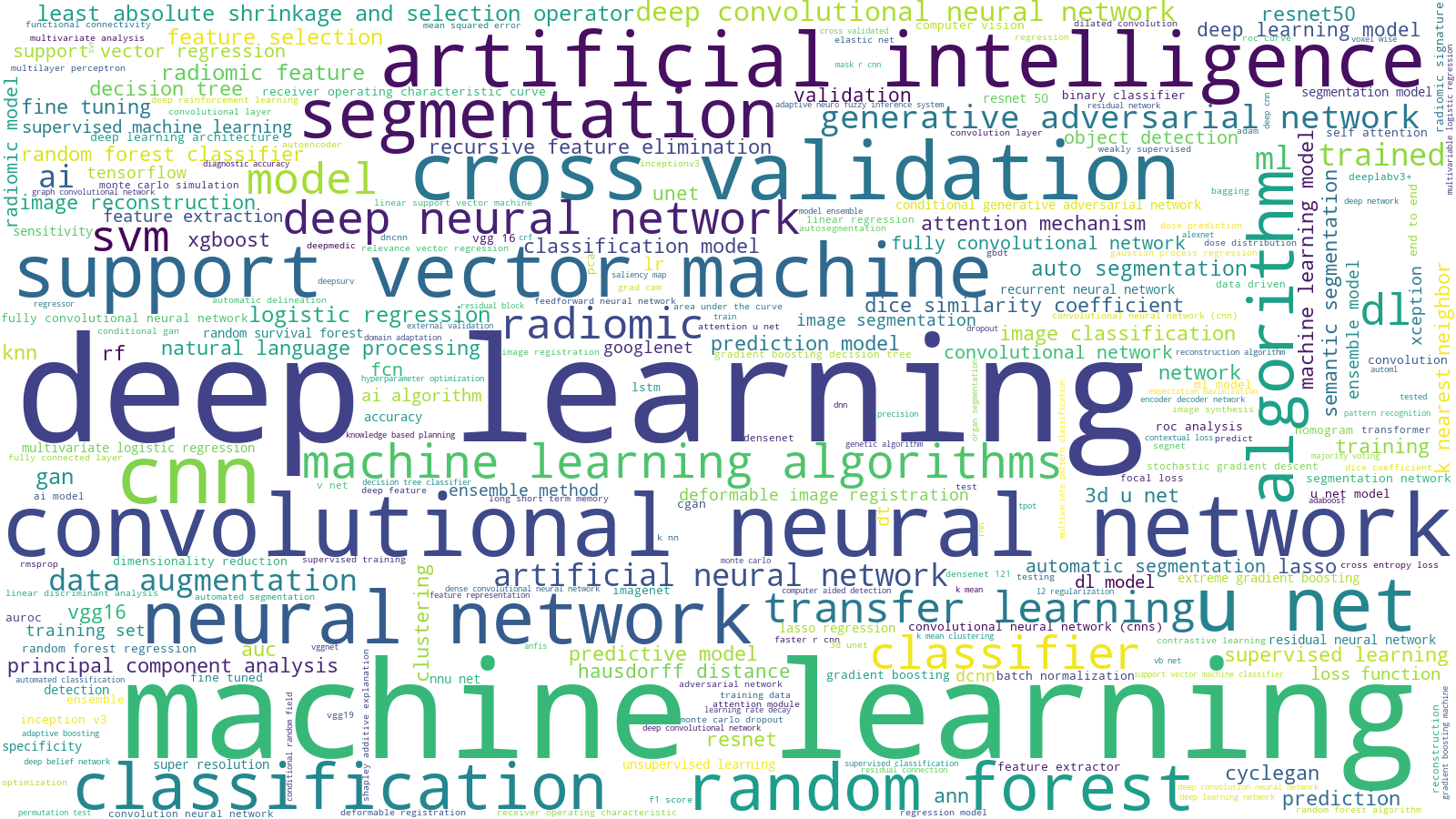}
  \includegraphics[width=0.15\textwidth]{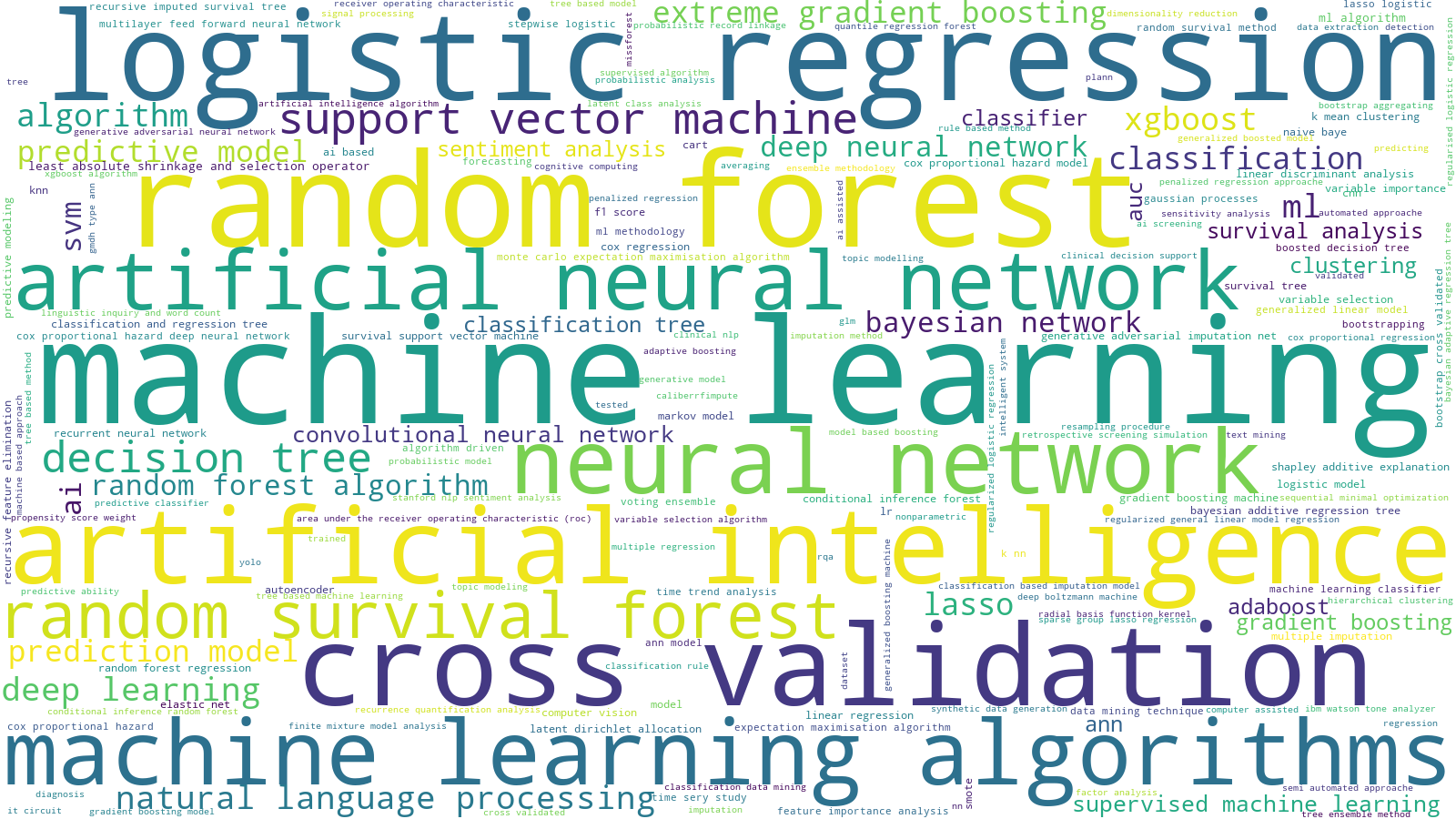}
  \caption{Subject category-specific word clouds of validated AI subfield keywords. 
  \textnormal{From left to right: (a) Math and Computational Biology, (b) Radiology and Imaging, and (c) Healthcare Science and Services. Word size reflects frequency of extracted AI concepts within each journal subset.}}
\label{fig:subfieldwordclouds}
\vspace{-0.3cm}
\end{figure}

To visualize the semantic patterns detected by our model, we generate word clouds from validated AI-related keywords (Figure~\ref{fig:subfieldwordclouds}). For each subject category, keywords from AI-relevant articles are aggregated and weighted by relative frequency, providing an overview of dominant methodological themes.
Clear differences emerge across disciplines. Math and Computational Biology journals feature a mix of classical machine learning and deep learning, with frequent mentions of \textit{neural networks} and \textit{random forests}. Mentions of \textit{convolutional} and \textit{graph neural networks} reflect applications to structured biological data. Radiology and Imaging is dominated by deep learning and computer-vision terms such as \textit{convolutional networks} (CNN), \textit{U-Nets}, and \textit{segmentation}, reflecting its focus on supervised image analysis, while tree-based and linear models appear far less often.
In contrast, Healthcare Science and Services emphasizes interpretable, clinically aligned models such as \textit{logistic regression} and \textit{random forests}, consistent with the needs of electronic health records and decision-support systems.
Across fields, ML and DL form shared foundations, but their form reflects domain-specific data: CNNs dominate imaging, statistical and representation learning shape biology, and interpretable models underpin healthcare. 
These trends should be viewed cautiously. We classify AI subfields using the ACM CCS system~\cite{rous2012ccs}, which, though widely used, is not tailored to biomedical research and may group diverse work into broad categories. Although supplemented with keyword extraction, we prioritized accurate category assignment over comprehensive coverage, so specialized methods may be underrepresented.
\subsection{Human Evaluation of LLM Annotations}
\begin{figure}[t]
\centering
\includegraphics[width=0.75\columnwidth]{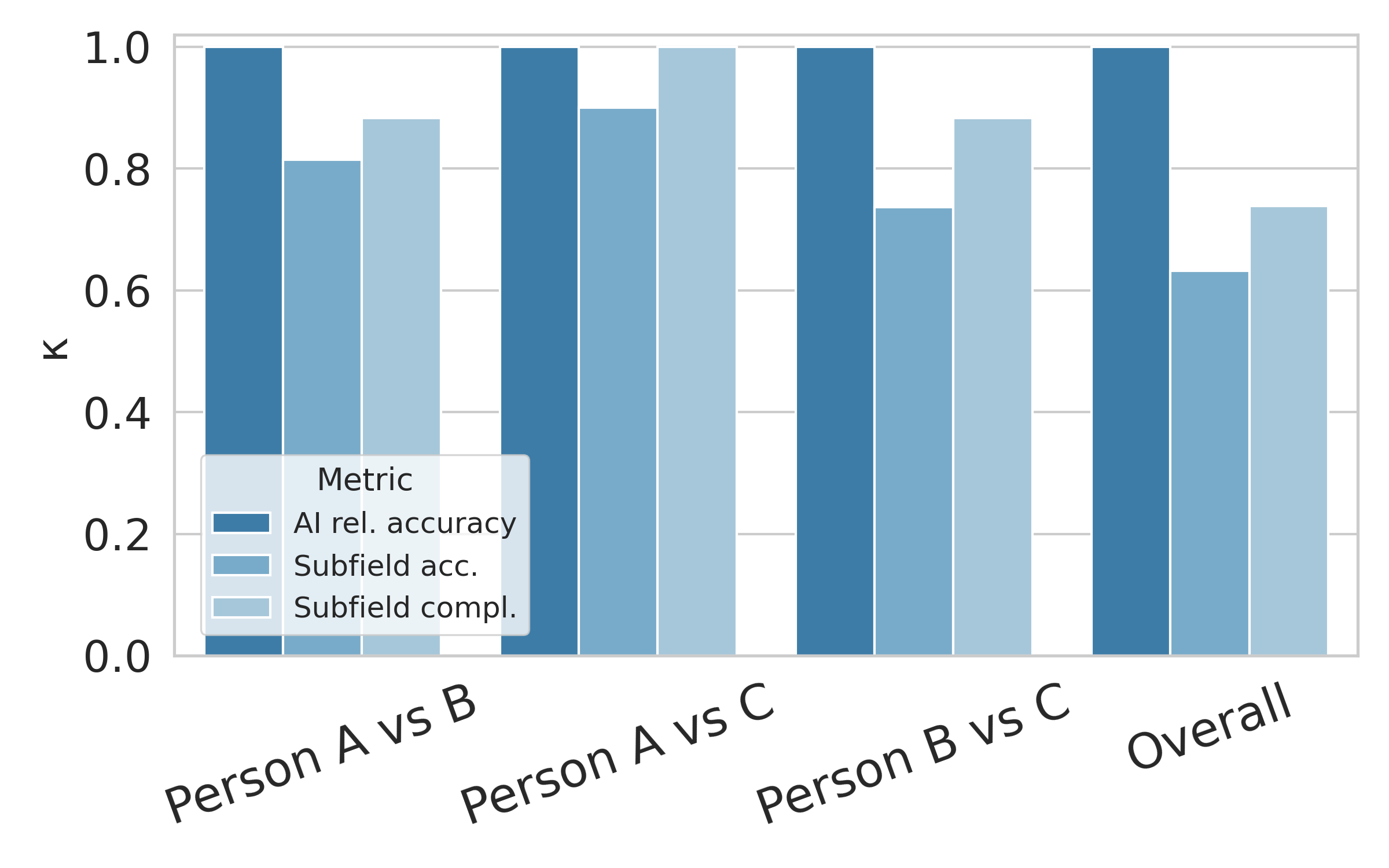}

\vspace{0.08cm}
{\small (a) Agreement ($\kappa$)}

\vspace{0.20cm}
\includegraphics[width=0.75\columnwidth]{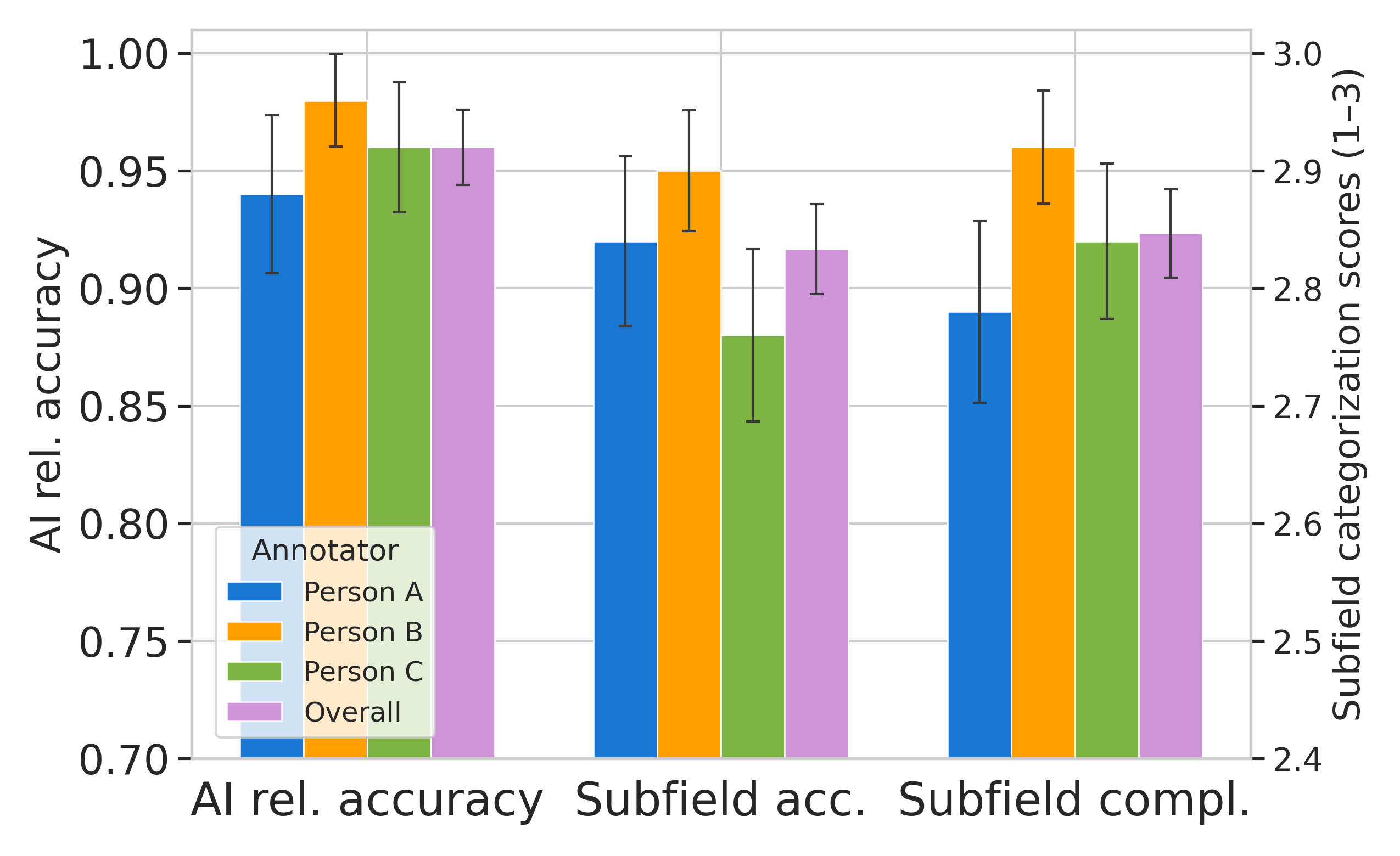}

\vspace{0.08cm}
{\small (b) Mean scores with Std.\ error}

\vspace{-0.1cm}
\caption{Human evaluation results. \textnormal{(a) Pairwise Cohen's $\kappa$ for each annotator pair and metric, with an overall bar showing Fleiss' $\kappa$ across all three annotators. (b) Per-annotator and overall mean scores with standard errors: AI-relevance accuracy (left axis, 0--1) and subfield accuracy/completeness (right axis, 1--3).}}
\label{fig:human-eval}
\vspace{-0.5cm}
\end{figure}
We evaluate the LLM pipeline on 100 stratified PMC-OA articles: 50 predicted AI-positive and 50 predicted AI-negative. Three author-team members (A, B, and C) participate in the study. We split articles into four sets of 25: one annotated by all raters and three assigned to a single rater, yielding 50 annotations per rater and 25 shared articles per pair. Each article is evaluated along three dimensions:

\vspace{0.05cm}
\begin{itemize}[leftmargin=10pt]
    \item \textbf{AI relevance accuracy}: whether the abstract involves artificial intelligence or machine learning.

    \item \textbf{Subfield accuracy (1-3)}: how accurately LLM-identified AI subfields or keywords reflect the content based on the ACM CCS system (1 = incorrect, 2 = partially correct, 3 = completely correct).

    \item \textbf{Subfield completeness (1-3)}: whether extracted AI subfields cover key AI-related technical aspects in the abstract based on ACM CCS system (1 = insufficient, 2 = partial, 3 = fully complete).
\end{itemize}

\vspace{0.05cm}

Annotators first evaluated AI relevance using only the title and abstract and were blinded to LLM relevance prediction. After submitting the relevance judgment, they were shown LLM-generated keywords and subfield assignments to evaluate subfield accuracy and completeness. We evaluate the reliability of LLM-generated annotations using $\kappa$ statistics, which account for agreement beyond chance. Pairwise agreement is measured using Cohen’s $\kappa$, while Fleiss’ $\kappa$ is reported for the 25 abstracts annotated by all three annotators. Since downstream subfield judgments are only meaningful when AI content is correctly detected, if the LLM misclassifies AI relevance, corresponding subfield accuracy and completeness scores are set to 1 by design. All Cohen’s $\kappa$ and Fleiss’ $\kappa$ statistics were computed on the 25 articles annotated by all three raters.

Figure~\ref{fig:human-eval}(a) shows that AI relevance annotation reaches substantial agreement across annotators ($\kappa > 0.8$), demonstrating consistent human judgments of LLM's AI relevance detection. Subfield accuracy and completeness show moderate to substantial agreement ($\kappa > 0.6$), indicating higher subjectivity in assessing technical precision and coverage while reflecting consistent interpretation. Figure~\ref{fig:human-eval}(b) reports individual and overall scores for each annotation dimension. AI relevance accuracy remains high across annotators, while slight variations in subfield completeness indicate stricter criteria among some annotators. Overall, the LLM-based pipeline demonstrates strong performance across evaluation dimensions, although author involvement may introduce confirmation bias.

\vspace{-0.05cm}
\section{Conclusion}
We introduced \textbf{BioMedJImpact}, a large-scale biomedical dataset built from over 1.7 million PMC-OA articles across 2,700 journals, integrating bibliometric, collaboration, and LLM-derived AI indicators to understand how biomedical publishing evolves in the AI era. Using this dataset, we examine how collaboration and AI engagement shape scientific impact. We show that greater collaboration intensity is associated with higher citation impact. We found that AI-related article share is positively associated with Impact Factor in BioMedJImpact-2019 but not in BioMedJImpact-2023 after multiple-testing correction. The LLM-based annotation pipeline is validated through human evaluation, confirming substantial agreement in AI relevance detection and consistent subfield classification. Overall, \textbf{BioMedJImpact} offers a comprehensive dataset at the intersection of biomedicine and AI and a scalable methodology for content-aware scientometric analysis.

\noindent\textbf{Ethical and Responsible Use.} This study analyzes biomedical articles from the PubMed Central Open Access Subset through PMC-authorized bulk-access services for non-commercial research. We only release journal-level aggregates and derived indicators, rather than full texts or patient-level records, and follow applicable article licenses and PMC terms. The resulting resource is intended for scientometric research only.
% \enlargethispage{3\baselineskip}

\bibliographystyle{ACM-Reference-Format}
\bibliography{references}

\end{document}